\documentclass{ws-brl}

\begin{document}

\markboth{Mayne, Adamatzky}
{Towards Hybrid Artificial-Slime Mould Devices}

%%%Publisher's Area%%%
\catchline{}{}{}{}{}
%%%

\title{Towards Hybrid Artificial-Slime Mould Devices}

\author{Richard Mayne}
\address{Unconventional Computing Group,\\
University of the West of England,\\
Frenchay Campus, Coldharbour Ln.,\\
Bristol, BS16 1QY\\
United Kingdom\\
richard2.mayne@live.uwe.ac.uk}

\author{Andrew Adamatzky}
\address{Unconventional Computing Group,\\
University of the West of England,\\
Frenchay Campus, Coldharbour Ln.,\\
Bristol, BS16 1QY\\
United Kingdom\\
andrew.adamatzky@uwe.ac.uk}

%%%fill in%%%
\maketitle
\begin{history}
\received{{\textbf D M Y}}
\revised{{\textbf D M Y}}
\end{history}
%%%

\begin{abstract}
The plasmodium of the slime mould \emph{Physarum polycephalum} has recently received significant attention for its value as a highly malleable amorphous computing substrate. In laboratory-based experiments, micro- and nanoscale artificial circuit components were introduced into the \emph{P. polycephalum} plasmdodium  to investigate the electrical properties and computational abilities of hybridised slime mould. It was found through a combination of imaging techniques and electrophysiological measurements that \emph{P. polycephalum} is able to internalise a range of electrically active nanoparticles, assemble them \emph{in vivo} and distribute them around the plasmodium. Hybridised plasmodium is able to form biomorphic mineralised networks, both inside the living plasmodium and the empty trails left in its wake by taxis, both of which facilitate the transmission of electricity. Hybridisation also alters the bioelectrical activity of the plasmodium and likely influences its information processing capabilities. It was concluded that hybridised slime mould is a suitable substrate for producing functional unconventional computing devices.
\end{abstract}


\section{Introduction}
\subsection{Slime mould as a computing substrate}
The myxomycete 'slime mould' {\emph Physarum polycephalum} is a motile, acellular amoeba which
possesses traits that may reasonably be considered to be 'intelligent'. For example, it can store information about its surroundings in an external spatial memory\cite{28}, calculate problems of computational geometry --- e.g. find the most efficient route between two points\cite{9} - and solve basic logic puzzles, such as navigating its way around a maze puzzle\cite{4,24}. At the molecular level, these intelligent functions are the product of manifold life processes such as chemotaxis, galvanotaxis\cite{1}, and photoreactions\cite{33}, which are effectively 'inputs' into the the slime mould. As such, {\emph P. polycephalum} can be considered as a biological amorphous computing device which we are able to program with inputs such as attractant/repellent chemical gradients and light\cite{3,5,11}, and whose outputs are the associated physical outcome of its computations, e.g. movement towards/away from a stimulus, the electrical responses of the organism. Significant progress has already been made in developing P. polycephalum into a functional unconventional computing device (or a 'Physarum machine', as dubbed in Ref. 2).

\begin{figure}
\centerline{\psfig{file=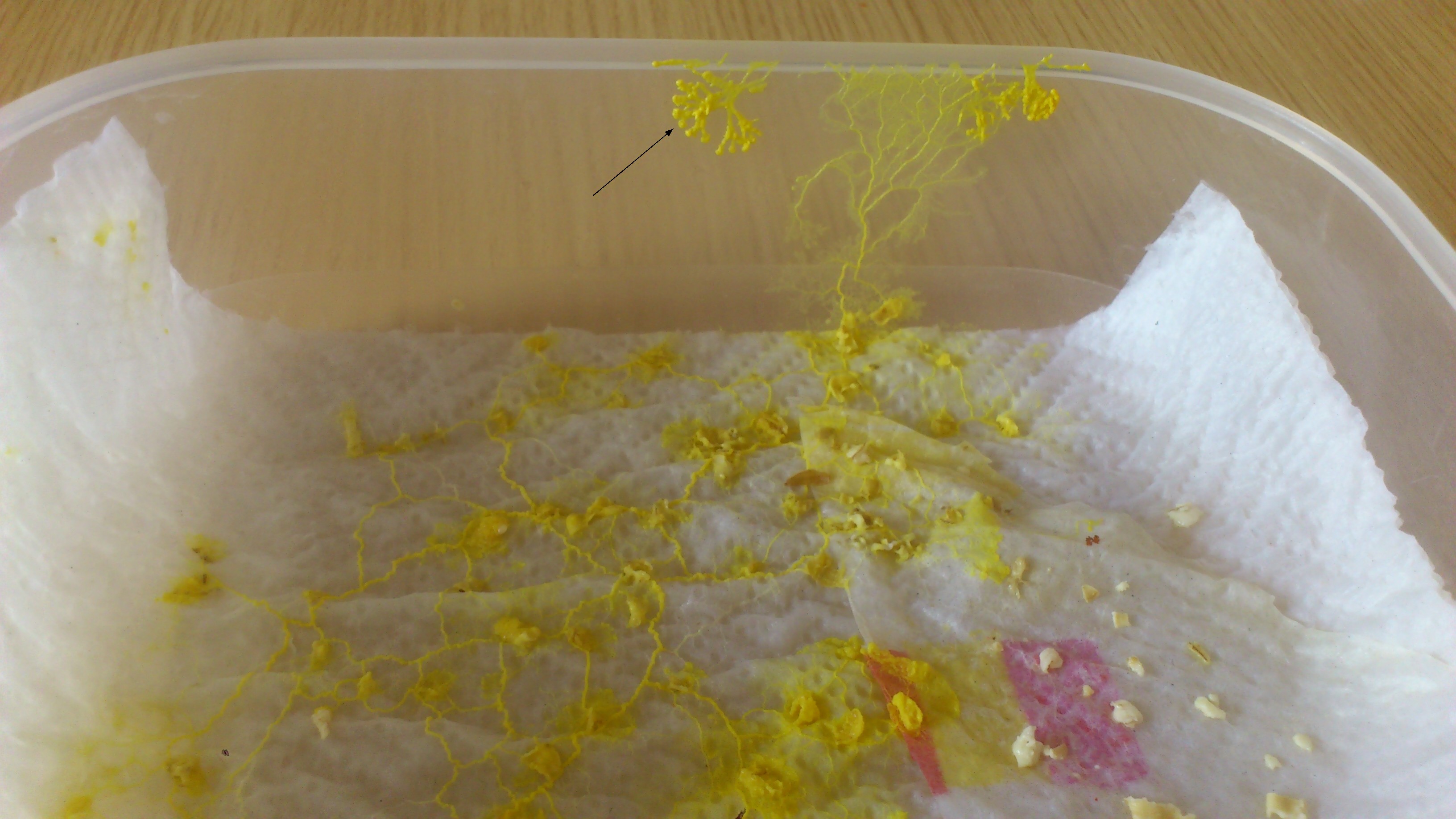,width=12.5cm}}
\vspace{8pt}
\caption{Slime mould {\emph P. polycephalum} growing on paper towels. Note the lattice structure of its plasmodial tubes connecting food sources (oats) and the 'fan shaped' morphology of the advancing anterior margin (arrowed).}
\end{figure}

The \emph{P. polycephalum} plasmodium (vegetative life-cycle stage, a multinucleate of a mass of protoplasm\cite{30}, \emph{pl.} plasmodia) is able to sense multiple chemical gradients of attractants and repellents in its surroundings simultaneously; it is therefore capable of massively-parallel processing\cite{3}, and this is the basis of its chemotaxis. It will propagate towards its desired nutrient source/sources by forming branching, interlinking tubes of plasmodium (Fig. 1) which gently oscillate via the activity of contractile proteins\cite{16}, providing motive force through instigating rhythmic flows in its protoplasmic core (cytoplasmic streaming). Once its advancing anterior margin (a fan-shaped arrangement composed of multiple converging protoplasmic tubes, see Fig. 1) finds its nutrient source –-- usually an oat flake in experimental situations –-- it is engulfed with pseudopodic invaginations in a manner similar to leukocyte phagocytosis\cite{7}. Once internalised in endocytotic vesicles, the internalised substrate is digested and mixed with the protoplasm, wherein the products are distributed throughout the organism via the free movement of fluids in its core (the endoplasm, see Fig. 2). In Ref. 1, Adamatzky demonstrated that the plasmodium will consume oat flakes soaked in coloured food dyes and mix them, thereby creating colour combinations; this illustrates how a Physarum machine may be manipulated to internalise a compound of interest, distribute it throughout its plasmodium and carry it to a location programmable with attractant/repellent gradients.

\begin{figure}
\centerline{\psfig{file=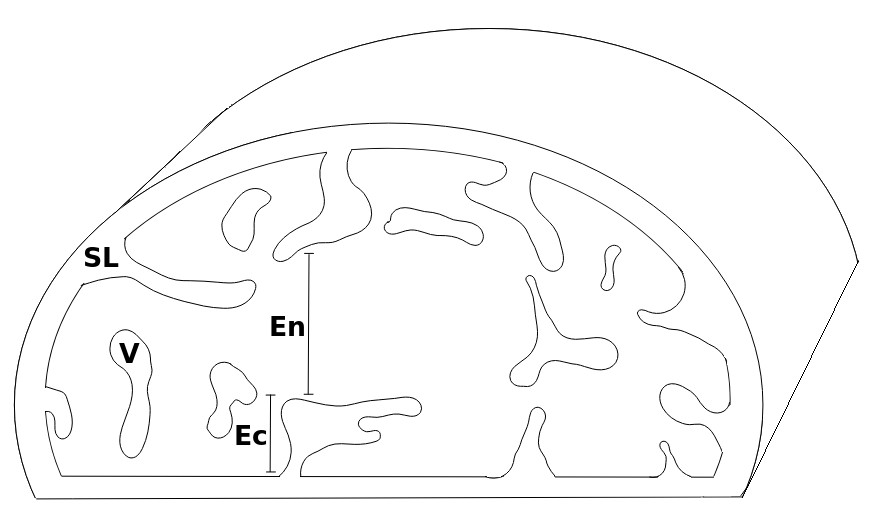,width=12.5cm}}
\vspace{8pt}
\caption{Schematic representation of a plasmodial tube in transverse section. SL = slime layer, En = endoplasm, Ec = ectoplasm, V = vacuole (see acknowledgements).}
\end{figure}

If developed sufficiently, the potential applications of biological computing technologies are
manifold, ranging from fully-integrated cybernetic augmentations/prostheses to eco-friendly
computers. An operational Physarum machine would be capable of massively-parallel sensing and be able to solve problems such as computation geometry and logic. Such a device would, however, necessarily require artificial components, e.g. for input/output interfacing. The aims of this investigation were therefore to conduct a scoping study into the construction of a hybrid Phyarum-artificial device, by introducing micro- and nano-scale circuit elements into the plasmodium.

There are three potential benefits in creating a hybrid slime mould-artificial device: the first is the generation of conductive pathways – i.e. wires – to facilitate the transmission of electrical information through the plasmodium. This could be achieved by coating the plasmodium in a layer of metal, infusing the interior with metallic particles, and/or using the plasmodium to deposit a solid wire in the hollow trails of effluent and dried slime left in the wake of propagation.

Secondly, previous studies have indicated that \emph{P. polycephalum} is able to assemble nano-scale metallic particles within the plasmodium into a range of structures\cite{22}. It may therefore be possible to generate discrete electrical devices within the plasmodium, which alter its information processing capabilities. If intraplasmodial assembly is indeed found to result in the generation of such devices, this technology could eventually be used to generate hybridised macro- scale devices, e.g. sensors.

Lastly, loading plasmodia with certain varieties of particle could allow us to manipulate their
behaviour with greater ease. If the electrical properties of a a hybridised circuit element alter in response to such an input, e.g. photoconductors, such devices could potentially alter plasmodial responses to the other inputs as a knock-on effect, thereby making its behaviour easier to manipulate in experimental conditions.

Any of these goals would contribute to the generation of a Physarum machine. The remit of
this scoping study was, therefore, to assess whether any of these goals may be possible to achieve and how future research should be directed.

\subsection{The electrical properties of  \emph{P. polycephalum}: a brief review}
As previously mentioned, electrical activity is one of the primary outputs of a Physarum
machine. {\emph P. polycephalum} both produces and responds to electrical events; this has been explored in depth by a number of authors, and some of their most relevant findings are summarised here.

In Ref. 12, Fingerle and Gradmann determined that the electrical potential of \emph{P. polycephalum} plasmodia exists, when measured at the membrane in a standard physiological buffer, in a depolarised state between -50 and -100mV. This potential is highly variable when exposed to varying solution pH and/or quantities of certain nutrients, such as ions and sugars: this suggests the existence of multiple transmembrane ion channels (i.e. H\textsuperscript{+} and K\textsuperscript{+} permeable proteins) which dictate – or at least contribute to – the electrical activity of the plasmodium. These results supported the earlier conclusions of Ridgway and Durham (see Ref. 29), who suggested that the movement of ions, specifically Ca\textsuperscript{2+}, across the plasma membrane of a moving plasmodium triggers movement. This was further supported by the more recent findings of Yoshiyama \emph{et al.} (see Ref. 34), who found that cytoplasmic streaming coincides with waves of Ca\textsuperscript{2+} influx. This is perhaps an unsurprising conclusion when one considers how animal muscles contract when exposed to the Ca\textsuperscript{2+} influxes triggered by neural input. From a computing perspective, the electrical output of the plasmodium can therefore be considered as directly proportional to the input from each transmembrane ion channel.

Furthermore, the electrical potential of a moving plasmodia oscillates, usually with an
amplitude of 1-15mV when measured as a differential voltage between the anterior and posterior
poles of the organism: this oscillation is rhythmic, ranging between one and two minutes per cycle and roughly correlates to the oscillatory patterns of cytoplasmic streaming\cite{20,32}. Oscillation of plasmodial electrical potential is presumed to be precipitated by the aforementioned Ca\textsuperscript{2+} influx patterns. Direct electrical stimulation of a plasmodium, however, does not alter the frequency or amplitude of the oscillations associated with cytoplasmic streaming, suggesting that electrical potential differences are not directly the cause of protein contraction, but rather an initiation event\cite{17,20}.

More recently, Adamatzky and Jones (Ref. 3) found that these patterns of oscillation are
relatively stable until environmental conditions alter significantly, wherein they revert to an
archetypal electrical pattern: for example, dehydration instigates a more irregular waveform which precedes a suitable response by the plasmodium, such as entering its resistant sclerotial life-cycle stage. This would suggest that plasmodial behaviour is intimately linked to electrical activity. Indeed, a range of tactile stimuli also cause changes in membrane potential, e.g. in response to being pierced with a needle or having a weight placed across it\cite{6,20}: these stimuli amount to sensory inputs, which in turn dictate the output, such as behaviour changes (i.e. negative thigmotaxis, sclerotinize) or activation/suppression of intracellular processes, e.g. repair mechanisms. These patterns of electrical response to stimuli all represent information processing in the processor that is the plasmodium.

But are these specific electrical properties the cause \emph{P. polycephalum's} intelligence? To answer this, it is necessary to compare \emph{P. polycephalum} plasmodia to animal neurons, the most researched-upon biological processors. Neurons carry electrical information in the form of action potentials via saltatory conduction, which are transmitted to other neurons at specialised junctions (synapses) through the actions of neurotransmitter chemicals. Neurotransmitters are released from the transmitting neuron in response to an action potential and are received by ligand-gated ion channels on the target, instigating a consequent depolarisation and hence propagating the action potential. A single neuron will be linked in this manner to \textgreater 10\textsuperscript{3} other neurons: the neuron is therefore a massively-parallel device (Beale and Jackson, 1990). Once an action potential has been delivered to its target --- such as an effector tissue or another neuron --- it will provoke a response, e.g. causing a muscle fibre to contract or a hormone to be released. The similarities between \emph{P. polycephalum} plasmodia and neurons are therefore evident: both are individual yet massively-parallel processors who produce electrical signals in their coordination of physiological responses.

The similarities between the two are further highlighted when considering the phenomenon
of memristivity in relation to memory. Memristors – literally 'memory resistors' – alter their
resistance proportionally with the magnitude of a unidirectional current flowing through it; when a current is removed, the device will 'remember' the last resistance it generated and re-assert it the next time current flows through it\cite{10,31}. Neurons are effectively biological memristors due to a phenomenon known as 'synaptic plasticity', wherein the excitability of a nerve is altered (i.e. potentiated or depressed) in response to frequent or infrequent use\cite{15}. Synaptic plasticity is thought to be the basis of learning and memory\cite{13}, and significant progress has been made in modelling the activity of nerves in circuits utilising memristors\cite{27}. The plasmodium of \emph{P. polycephalum} has also recently been found to be memristive, thereby providing a potential electrical basis for its learning abilities and a mechanism for modelling its activity with circuits\cite{14,26}. Exhibiting electrical properties which are integral to biological computing further emphasises the value of a \emph{P. polycephalum} computing device, and provides us with insights into how we may further manipulate its electrical properties to our own advantage.

\subsection{Hypothesis and proposal}
\emph{P. polycephalum} is an intelligent organism that is highly analagous to the most powerful computer known to man – the human brain – yet it is easy to culture, tolerant to a variety of analytical techniques and has no associated ethical issues surrounding its use as a research organism\cite{18}. It would appear that, like neurons, the behaviour of \emph{P. polycephalum} is both reflected in and dictated by electrical events. We propose, therefore, that through the manipulation of \emph{P. polycephalum's} electrical properties, and hence the manual adjustment of its computing abilities, the objectives described in section 1.1 can best be achieved. As such, both the input and the output of a Physarum machine may be designed, and any artificial components added would augment or/and modify either of these.

In Mayne \emph{et al.} (see Ref. 22), it was ascertained that \emph{P. polycephalum} can internalise suspensions of starch-coated magnetite multi-core nanoparticles (NPs) dropped onto homogenised colonies, via endocytosis. It stores a certain quantity of the suspension in a variety of sizes/structures within the protoplasm whilst depositing the rest in a trail left in the wake of its movement. In the current study, plasmodia were treated with the following electrically active NPs:

\begin{romanlist}
\item Zinc Oxide (ZnO), 100nm\textsuperscript{\o} (Sigma-Aldrich, Gillingham, UK), dispersed in water. These
were chosen for their being both semiconductors and photoconductors\cite{23}, giving them the potential to form a range of devices and also alter \emph{P. polycephalum's} response to light.
\item Multi-core magnetic magnetite ($Fe_3O_4$), 200nm\textsuperscript{\o} (ChemiCell GmbH, Berlin, Germany),
dispersed in water. These were chosen for their being electrically conductive, magnetic (or
more accurately, superparamagnetic) and memristive when in nano-scale arrays\cite{19}; the integration of such NPs could potentially facilitate the generation of miniaturised
electronic components, solid state wires and the magnetic storage of information.
\item Tungsten (VI) oxide ($WO_3$), 100nm\textsuperscript{\o} (Sigma-Aldrich, Gilingham, UK), dispersed in water.
Also semiconductors, this variety of NP has a different bandgap to zinc oxide; they are also
thought to be less toxic.
\end{romanlist}

The study was conducted by treating plasmodia with various quantities of these NPs and thoroughly characterising the alterations --- if any --- in the electrical properties of the organism. The patterns of NP uptake and intraplasmodial assembly were comprehensively investigated with a range of imaging techniques, namely scanning electron microscopy (SEM), transmission electron microscopy (TEM), energy dispersive X-ray microanalysis (EDX) and confocal microscopy.

\section{Methodology}
\subsection{Slime mould culture}
Stock slime mould was cultured in plastic boxes on moistened kitchen towel in the dark at
room temperature, and was fed with porridge oats. Cultures were periodically transferred to fresh boxes/kitchen towel as they outgrew their environment. Cuttings of colonised kitchen towel were transferred to 2\% nutrient agar plates in order to grow large amounts of metabolically active plasmodium for experimental use. Colonies receiving treatment with NPs were homogenised with a scalpel blade and inoculated into the centre of 9cm Petri dishes filled with 2\% non-nutrient agar.

Plasmodia treated with NPs had varying quantities of relevant NP suspension dropped on
them with a pipette. Thin moats were cut in the agar around the colonies to prevent diffusion of NP suspensions through the agar (see Ref. 2). The colonies were then left to proliferate and migrate towards oats placed in the peripheral regions of their Petri dishes for 48h (Fig. 3).

\begin{figure}
\centerline{\psfig{file=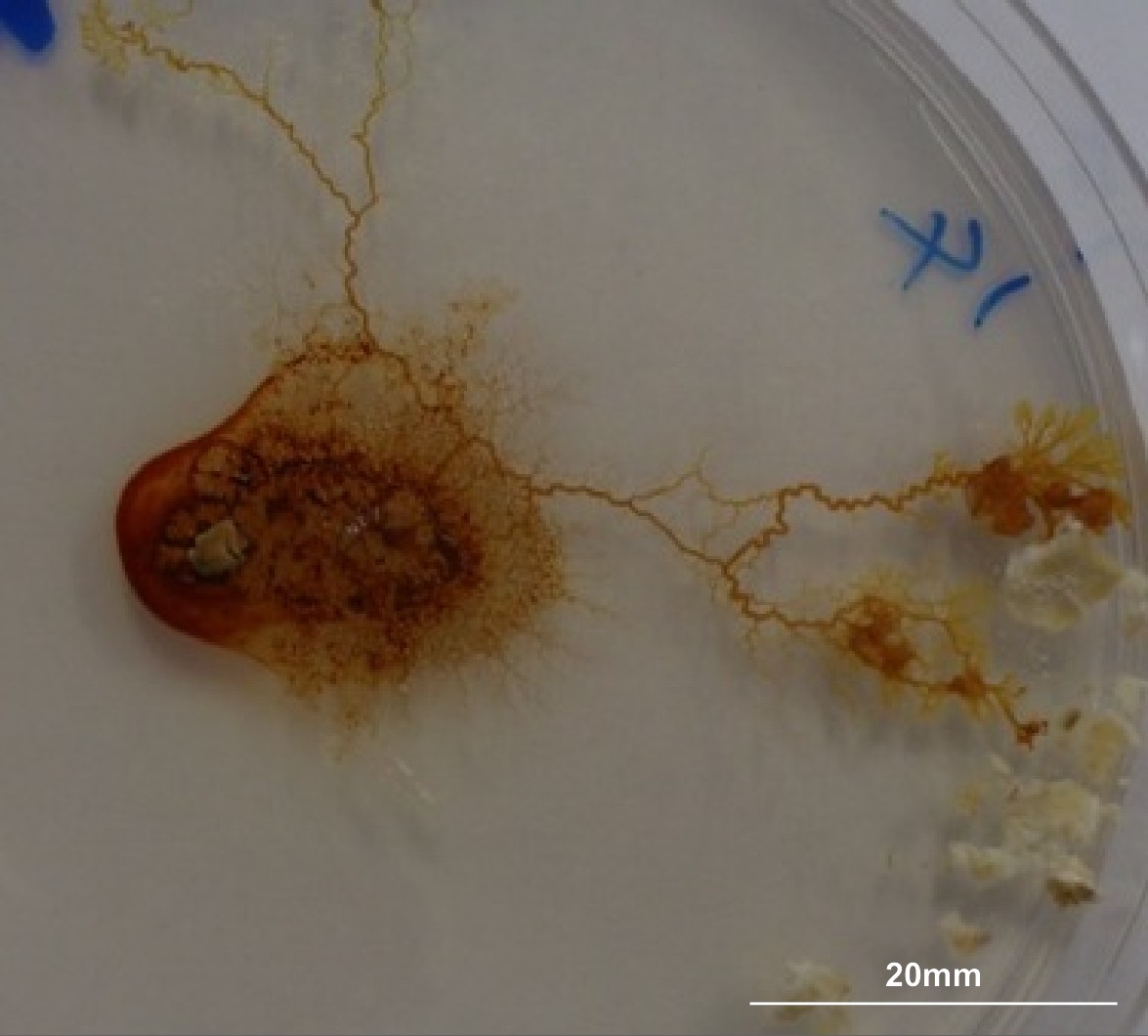,width=10cm}}
\vspace{8pt}
\caption{Pladmodium loaded with magnetite NPs migrating towards oat flakes.}
\end{figure}

A range of preliminary experiments were conducted to determine the highest possible
concentration of NPs the slime mould cultures could absorb without observable detrimental effects: this was achieved by exposing colonies to serial dilutions of NP suspensions. The most appropriate suspension concentration was then used in all further experiments.

\subsection{Electrical measurement}
Colonised oats were removed from their Petri dishes and placed on a blob of 2\% non-
nutrient agar(c. 1.5ml) overlying a strip of aluminium tape (190x8mm) stuck to a glass microscope slide. Another agar blob and aluminium tape device was present 10mm away from this, which had a fresh oat placed on it to encourage the slime mould to grow from one blob to another, thereby creating a 'wire' between the two electrodes (Fig. 4). Once a slime mould 'wire' had grown (24--48 hours), the exposed ends of aluminium tape were connected to test leads and the differential voltage across the plasmodium was measured using a Picotech ADC-24 data logger (Pico, UK).

\begin{figure}
\centerline{\psfig{file=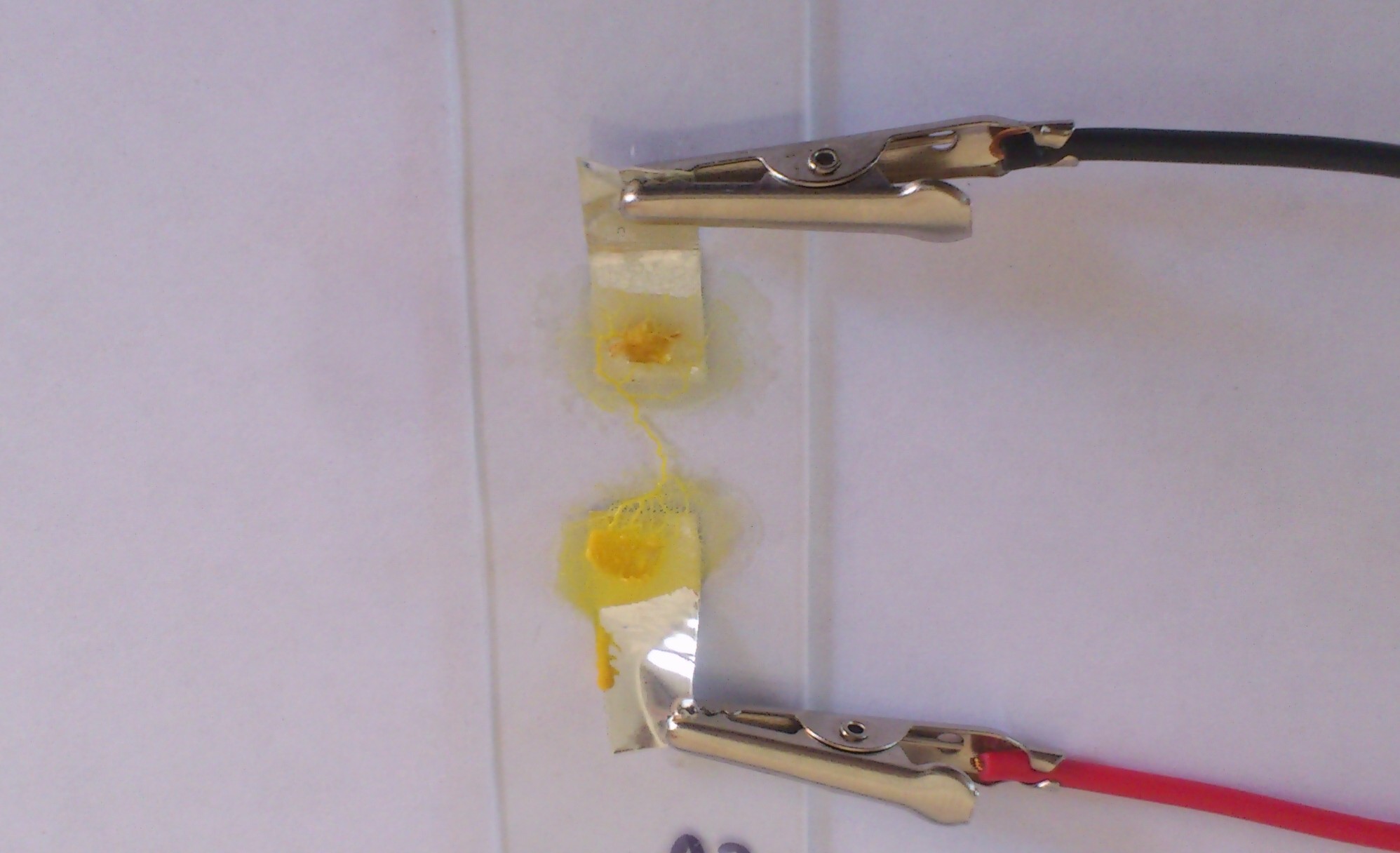,width=13cm}}
\vspace{8pt}
\caption{A control slime mould colony linking two agar blobs overlying electrodes with a single plasmodial tube (tape shortened for illustrative purposes).}
\end{figure}

It is important to note that in these circuits, agar blobs must be considered as two resistors – i.e. the electrical properties being measured do not reflect the 'true' voltage/resistance of the plasmodium. In preparatory studies, the resistance of single agar blobs was measured with a Picotech PT-104 data logger. It was noted that their resistance increased gradually (from an average initial value of 18.4K$\Omega$) over time as the blobs desiccated. Measures were consequently taken to prevent desiccation of blobs during measurement and therefore reduce variation between samples: the test environments (Petri dishes) were sealed in the presence of a small quantity of water to maintain humidity, and longer experiments had small quantities of deionised water dispensed onto the blobs to further prevent desiccation, depending on the ambient temperature.

A range of variables were measured by hand from the resulting traces, including the average
wavelength and average amplitude of oscillation, as well as the average voltage at which the
plasmodium oscillated. This range of measurements was chosen to give as thorough a
characterisation of the normal oscillatory activity as possible for the prescribed experimental setup. Plasmodia that did not oscillate were discounted from the investigation.

The resistance of plasmodial tubes was also measured using an Agilent U1253B multimeter.
In oscillating plasmodia, the resistance was measured over the course of an entire oscillation and averaged. Measurements were taken from intact oscillating tubes as well as the desiccated 'empty' tubes left behind following plasmodial migration or death.

\subsection{Confocal microscopy}
Plasmodia were inoculated onto glass slides coated with a 1mm layer of 2\% low-
fluorescence agar before being treated with fluorescent NPs (fig. 5). Slides were then left in the dark for 24 hours to proliferate and migrate towards a supplied food source. A narrow moat was cut in the agar around the inoculation point to prevent the diffusion of fluorescent particles through the agar: this region of agar was removed from the slide prior to imaging. The two varieties of NPs used were:

\begin{romanlist}
\item Magnetite multi-core ferromagnetic NPs conjugated to a fluorophore (perylene) (Chemicell GmBH, Germany), 200nm\textsuperscript{\o}. Aside from the fluorescent label, these particles were identical to the magnetite NPs used in other experiments.
\item FluoSpheres\textsuperscript{\textregistered} carboxylate-modified latex nanospheres (Life Technologies, UK), 100nm\textsuperscript{\o}. These were chosen by virtue of being 100\% bio-compatible, highly fluorescent and having carboxylate coupling surfaces, which could potentially facilitate nano-assembly in situ.
\end{romanlist}

\begin{figure}
\centerline{\psfig{file=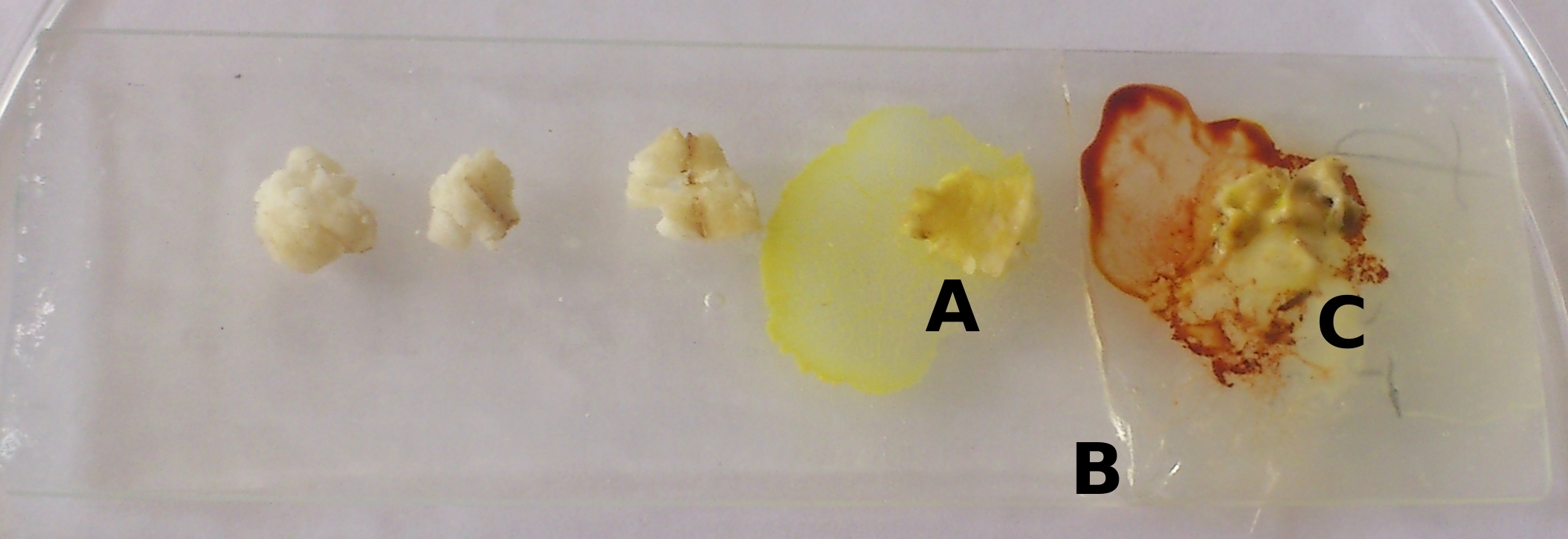,width=13cm}}
\vspace{8pt}
\caption{Setup for confocal microscopy. Note how the slime mould (A) has migrated across the moat (B) towards the supplied oat flakes, but that the excess nanoparticle suspension at the inoculation point (C) has not diffused across.}
\end{figure}

A fluorescent nuclear dye, Hoescht 33342 (Life Technologies, UK), was also used. Slides
were then washed in pH 7.1 potassium phosphate buffer, before being visualised with a Perkin
Elmer Ultraview FRET H spinning disk confocal microscope. Samples were analysed live, i.e.
without fixation or permeabalisation, in order to observe the movements of internalised NPs in real-time.

\subsection{Electron microscopy and microanalysis}
Samples of plasmodial tubes of roughly 1mm\textsuperscript{3}, including their agar base, were taken and prepared for electron microscopy. Scanning electron microscopy (SEM) samples were mounted on specimen stubs with adhesive carbon tabs and visualised in a Philips XL-30 environmental SEM: the apparatus was operated in 'wet' mode at pressures between 4.5 and 6.5 Torr, on a Peltier cooling stage at 5$^\circ$C. Semi-thin resin sections of plasmodium (see below) mounted on copper specimen grids were also analysed with SEM. Energy dispersive X-ray spectroscopy (EDX) was performed upon all SEM samples, using an Oxford Instruments Link system. All EDX analyses were performed in high vacuum in order to maximise resolution.

Tube sections were also fixed and embedded in resin for SEM and transmission electron
microscopy (TEM). Samples received initial fixation with 2\% glutaraldehyde in pH 7.1 potassium phosphate buffer for approximately 1 hour at room temperature, before being rinsed in the same buffer three times. They were then refrigerated overnight.

It was found that agar-based samples are almost invisible when submerged in embedding
resin which resulted in significant tissue loss: in an attempt to remedy this, 1ml of 1\% methylene blue solution was added to the buffer solution prior to refrigeration to stain the agar blocks. Samples then underwent post-fixation in 1\% aqueous osmium tetroxide for 1 hour, before being dehydrated in graduated concentrations of ethanol. Samples were embedded in Taab embedding resin, before being cut on a Reichert-Jung Ultracut E ultramicrotome. Semi-thin sections were prepared for SEM and light microscopy and ultra-thin sections were stained with uranyl acetate (5 minutes) and lead citrate (20 minutes) for imaging in a Philips CM-10 TEM.

\section{Results}
\subsection{Notes on the culture of altered plasmodia}
The \emph{P. Polycephalum} plasmodium appeared to be highly tolerant to treatment with magnetite and tungsten oxide NPs, but was considerably less so when treated with zinc oxide. Experimental colonies treated with undiluted stock solutions of magnetite survived reasonably well; zinc oxide, however, killed 100\% of colonies at concentrations above 6.25\% w/v. Tungsten oxide did not appear to be biocidal at concentrations of 25\% w/v. The morphology of colonies treated with tungsten oxide and zinc oxide NPs was noticeably erratic, whereas those treated with magnetite appeared to be relatively normal, albeit slightly emaciated. It was also noted that magnetite NP-treated colonies had a greater tendency to migrate in random directions rather than towards the supplied oat flakes.

Examination of intact plasmodia treated with magnetite NPs with light microscopy revealed
the presence of large brown/black deposits (Fig. 6 (A)), which could be seen being transported
through the endoplasm; these were putatively identified as large aggregates of magnetite. Static, but otherwise similar deposits were also observed in significant quantities in the peripheral regions of active tubes and spanning the whole diameter of desiccated, 'empty' tubes. No discrete NP deposits were easily visible in plasmodia treated with zinc oxide or tungsten oxide NPs, indicating that if any metals were present, they were likely to still be nano-scale and hence invisible via light microscopy. The empty tubes of plasmodia treated with tungsten oxide NPs, however, appeared darker than those of their unaltered equivalents (Fig. 6 (B)).

\begin{figure}
\centerline{\psfig{file=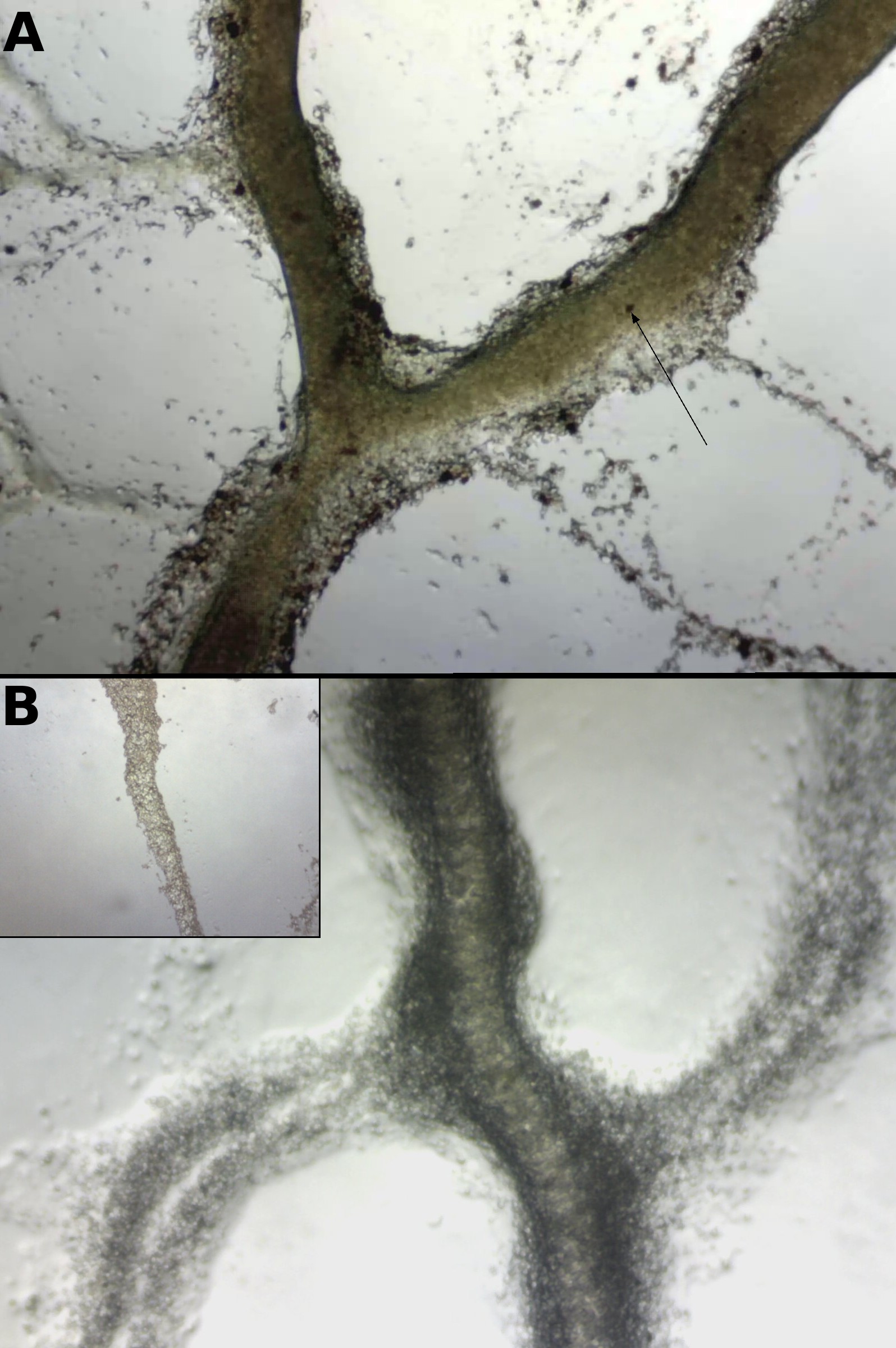,width=13cm}}
\vspace{8pt}
\caption{Light microscopic image plate. (\textbf{A}) A minor plasmodial tube loaded with magnetite nanoparticles. Note the large black depisits which were identified as aggregated magnetite. The arrowed deposit was observed to move down the tube over time. (\textbf{B}) An empty tube from a plasmodium treated with tungsten oxide nanoparticles. Note the darker shadow about the tubes in comparison the inset image of an unaltered empty tube.}
\end{figure}

\subsection{Confocal microscopy}
Confocal microscopy revealed that test plasmodia were successfully loaded with both
varieties of NP used (fluorescent magnetite NPs and latex nanospheres), wherein they were
dispersed thoughout the plasmodium via fluid movements in the endoplasm. NPs were observed
depositing to the peripheral, non-motile regions of the plasmodium, i.e. the ectoplasm and slime layer (see Fig. 2), in a variety of different sizes and structures. When administered in isolation, magnetite was found to form aggregative structures of a variety of sizes (Fig. 7 (B)). When latex nanospheres were applied with magnetite NPs, complexes with fluorescence returns for both varieties of particle were observed. These complexes ranged in size from objects no bigger than a single latex nanosphere with a similarly sized quantity of magnetite (Fig. 7 (C)), to much larger (over 1$\mu$m), presumably multiple-nanosphere structures (Fig. 7 (D)). Whilst it is worth noting that any measurements made with this technique are subject to a certain degree of inaccuracy due to a) the depth of field of the confocal effect (optical sectioning), and b) the different comparative brightnesses of the particles used (the latex beads were significantly more fluorescent than magnetite), it is nevertheless apparent that multi-NP structures were present whose sizes were significantly larger than their monomeric equivalents.

\begin{figure}
\centerline{\psfig{file=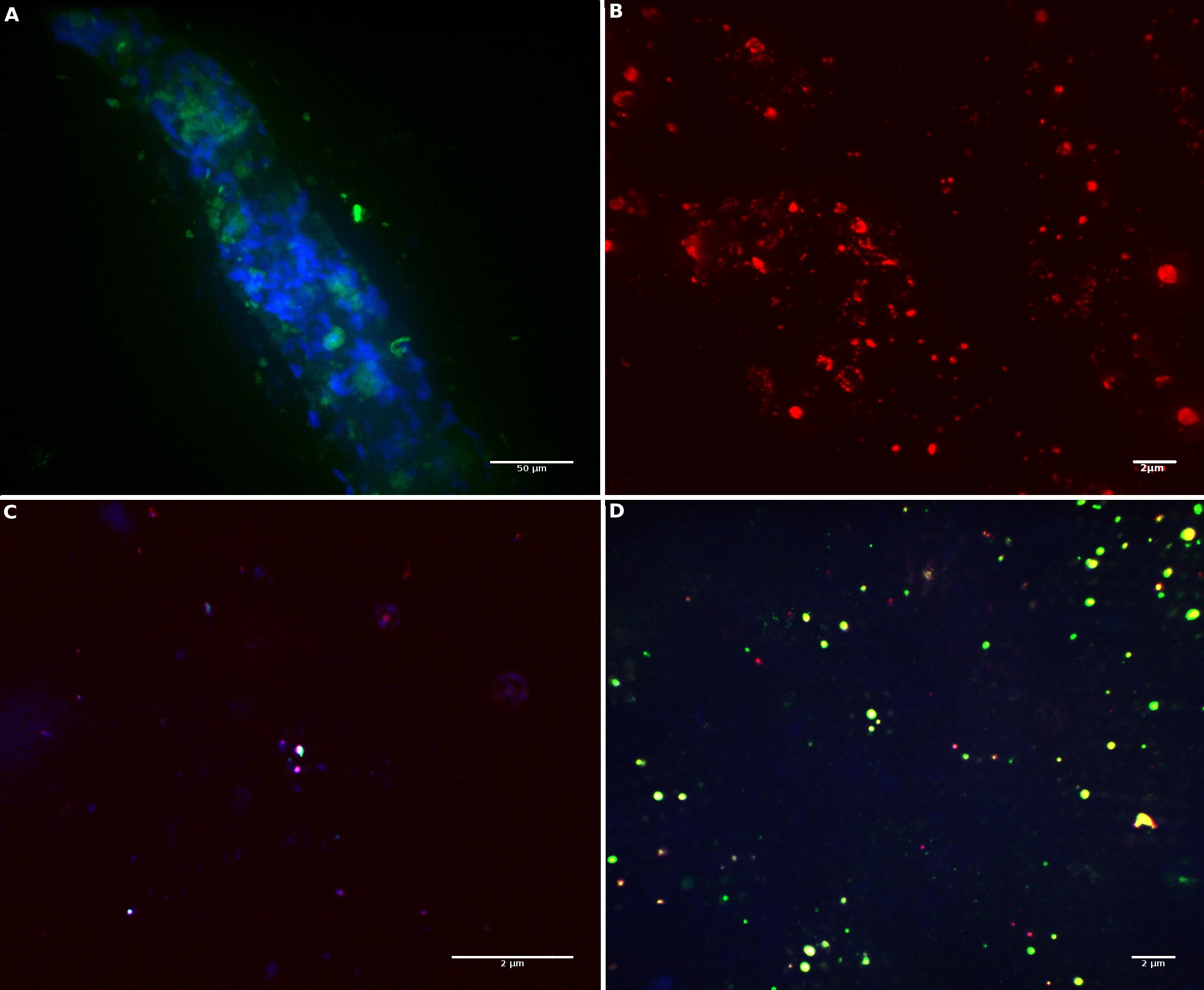,width=13cm}}
\vspace{8pt}
\caption{Confocal microscopy image plate. (\textbf{A}) Low power, minor plasmodial tube of a control
colony showing nuclei (blue) and the general tube structure (green) (\textbf{B}) Magnetite deposits (red) stuck in the peripheral regions of a major plasmodial tube, forming a variety of different structures. (\textbf{C}) A multi-particle structure (central), containing a latex bead (green) and magnetite (red and blue). Due to its size, this strucutre is likely to be a single latex bead bound to a similary sized quantity of magnetite NPs. (\textbf{D}) Image taken from the peripheral region of a major plasmodial tube, showing many NP deposits, some of which appear to contain both varieties of NP (as denoted by their fluorescence returns and relative sizes).}
\end{figure}

Confocal video footage of live plasmodia showed real-time transport of fluorescent particles
through the endoplasm (Fig. 8). The peripheral regions of the tubes (the boundaries between
endoplasm, ectoplasm and slime layer being indistinct with this technique) appeared to be partially permeable to the contents of the endoplasm, through which the fluorescent particles would move more slowly, with their speed of movement being inversely proportional to their size and the depth of infiltration into the periphery. The deposition of particles to the external layers is analagous to the action of a sieve, i.e. all contents too large to fit between the 'gaps' (an effect presumably created by the high density of vesicles in the ectoplasm and reduced fluid movement) became stuck. The mobile contents of the endoplasm contained a higher proportion of particles of all sizes, whereas the periphery appeared to contain more large deposits than small; this observation supports the aforementiond sieve hypothesis.

\begin{figure}
\centerline{\psfig{file=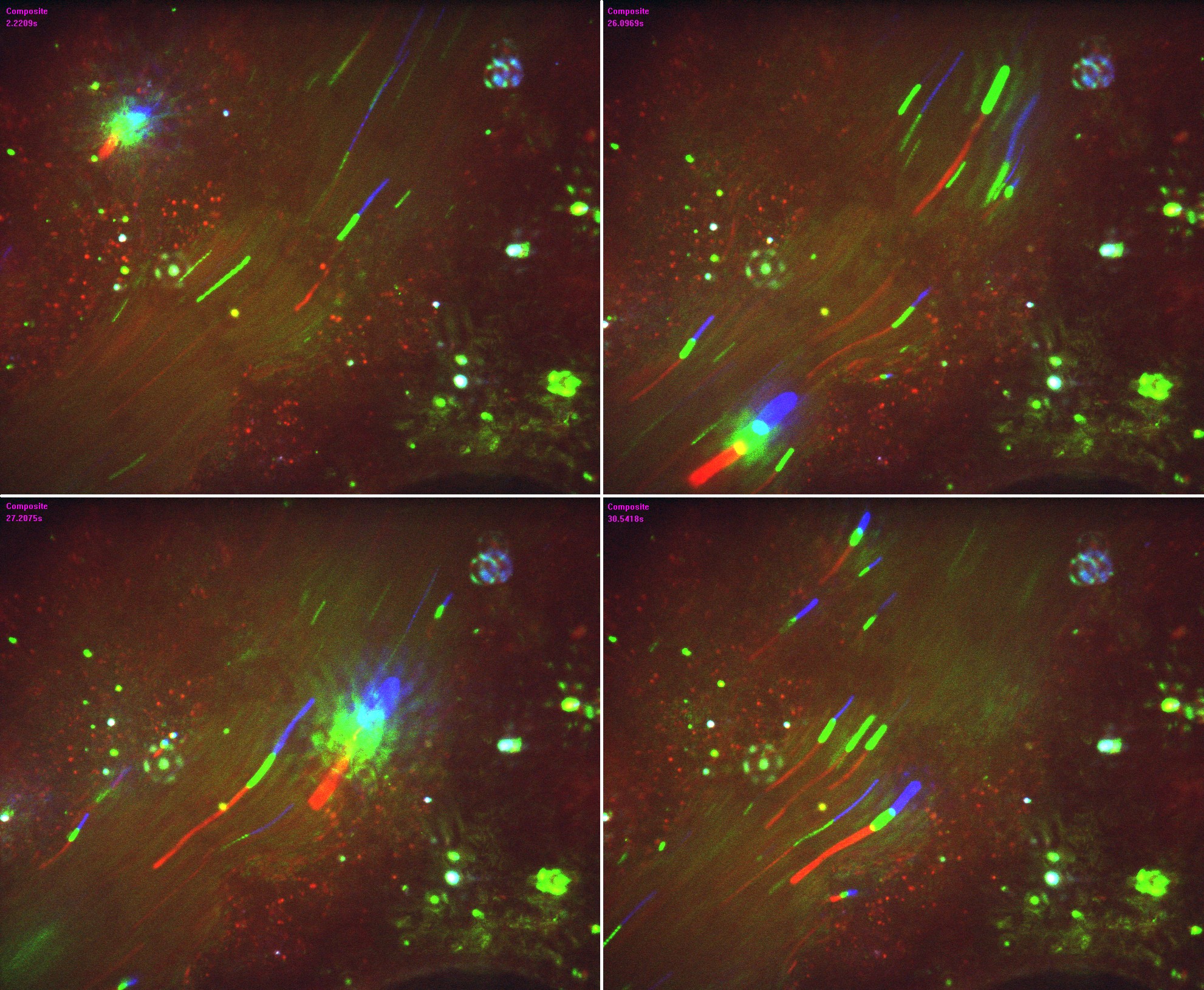,width=13cm}}
\vspace{8pt}
\caption{Screenshot gallery from a video of a major plasmodial tube transporting a significant quantity of latex beads (green) and magnetite (red; blue and red for larger deposits); the motile fluid core (running diagonally from bottom left to top right) is transporting a significant quantity of particles, many of which become trapped in the static peripheral regions (lower right).}
\end{figure}

\subsection{Electron microscopy and microanalysis}
No zinc was detected in any experimental colonies with SEM/EDX. This was hypothesised
to result from a) Physarum intentionally not absorbing/actively removing zinc, possibly due to its apparent intolerance to it, and/or b) the incredibly low concentration of zinc that had to be used, again due to the intolerance observed. Consequent electrical testing upon colonies loaded with zinc oxide was therefore not performed, as it was reasoned that even if minute amounts were inside the colony, their concentrations would be insufficient to alter plasmodial electrical or/and behavioural properties.

Both iron and tungsten were detected in altered plasmodia treated with relevant NPs. In
transverse section, a variety of structures ranging in size from 0.2--4.8$\mu$m were found which were revealed to contain both iron and tungsten in differing percentages (Fig. 9). The frequency and distribution of deposits per tissue section was highly variable, although it was more common for larger deposits to be held in vacuoles and situated within the ectoplasm, with \emph{vice versa} being true for endoplasmic deposits, although very few discrete endoplasmic deposits of metals were found. A broader-beam analysis of the endoplasm, however, revealed trace amounts of iron and tungsten (data not included).

\begin{figure}
\centerline{\psfig{file=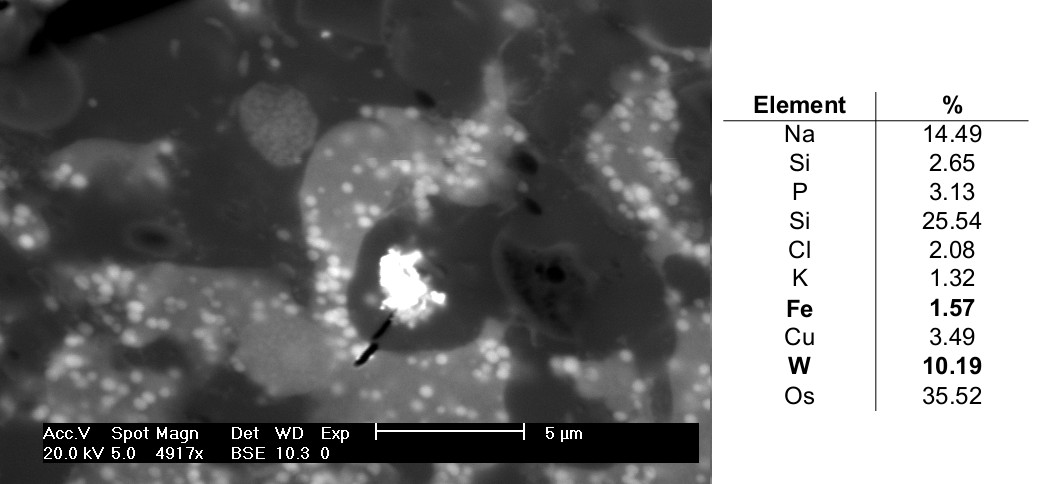,width=13cm}}
\vspace{8pt}
\caption{SEM image from a plasmodium treated with magnetite and tungsten oxide nanoparticles in transverse section. The results of an EDX elemental content analysis performed upon the large white polygon (indicated by a black cross) is included (carbon and oxygen omitted). Note that sulphur, copper and osmium are residues from the fixation/embedding process.}
\end{figure}

Viewing intact plasmodial sections from a 'top down' perspective revealed little useful
information, save for thin skeins of (EDX-confirmed) iron being visible on the outer surfaces of plasmodia treated with magnetite NPs (data not shown).

NP-containing deposits and were identified in TEM sections by a) the morphological
similarities they shared with SEM/EDX-identified deposits, b) the absence of similar structures in control sections and c) the manufacturer-supplied size ranges for each particle variety. The structures of these deposits (Fig. 10) were highly variable in a manner consistent with confocal and SEM observations. Tungsten oxide NP-containing deposits were formed from angular monomers between 50 and 100nm; their appearance was consistent with published images of similar NPs\cite{25}. Magnetite deposits had a dense, aggregative appearance continuous with their fluorescent equivalents (see section 3.2). All NP deposits were extremely electron-dense and tended to overlap, which obscured their fine structure. As indicated by confocal and SEM/EDX findings, some deposits were observed to combine the structural characteristics of both NP varieties. The sizes of these multi-NP structures ranged from ~10nm to \textgreater 1$\mu$m and were distributed in a manner consistent with the SEM observations, i.e. larger and more frequent towards the peripheral regions, with larger deposits having the propensity encapsulated in vesicles. NP-containing strucutres were also identified outside the outer membranes (Fig. 10, lower).

\begin{figure}
\centerline{\psfig{file=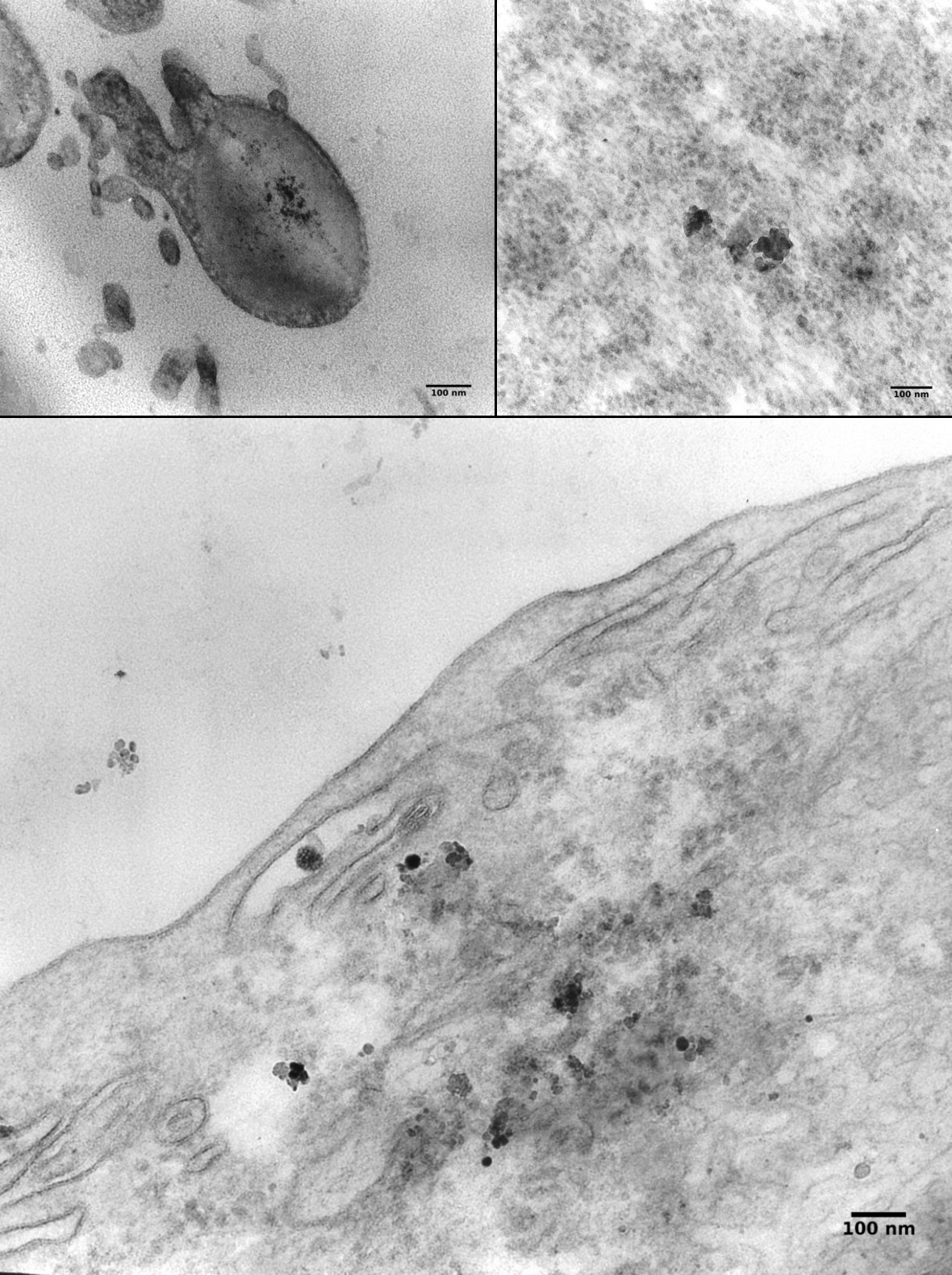,width=13cm}}
\vspace{8pt}
\caption{TEM plate. (\textbf{Top left}) Dense, granular structures within a slime vesicle whose appearance is consistent with that of magnetite. (\textbf{Top right}) Angular, crystalline-appearing deposits continuous with protoplasm, putatively identified as tungsten oxide. (\textbf{Lower}) A mixture of magnetite and tungsten structures continuous with the protoplasm and in vesicles at the outer membrane (low contrast). Note the presence of a putative deposit outside the tube.}
\end{figure}

\subsection{The electrical properties of altered plasmodia}
\subsubsection{Notes on experimental procedure}
It was noted during initial investigations that oats colonised with altered plasmodia had an
unacceptably high rate of failing to propagate: the methodology was consequently altered by
transferring a set quantity of homogenised unaltered plasmodium onto the agar blobs before a set amount of NP suspension was dropped on top. This resulted in some contamination of the agar
blobs with NP suspension, but did not appear to significantly alter the electrical properties of the agar when individual contaminated blobs \emph{sans} plasmodium were analysed.

Measurement of electrical potential in control colonies produced results consistent with
those of Adamatzky and Jones (see Ref. 3), in that a rapid depolarisation (peaking between -5 and -10mv) is observed at the point of inital colonisation of the second electrode, followed by repolarisation leading in to a 'resting' phase, where the potential difference oscillated somewhere between -3 and 3mv (Fig. 11) until sclerotinization or migration away from the electrode. Oscillations were seen during repolarisation and resting phases (Fig. 11), as were wave packets. It should also be noted that the polarity of the plasmodium was interchangeable, i.e. sometimes they would depolarise and oscillate at a negative potential difference, whereas other times they would hyperpolarise and oscillate at a positive potential difference.

\begin{figure}
\centerline{\psfig{file=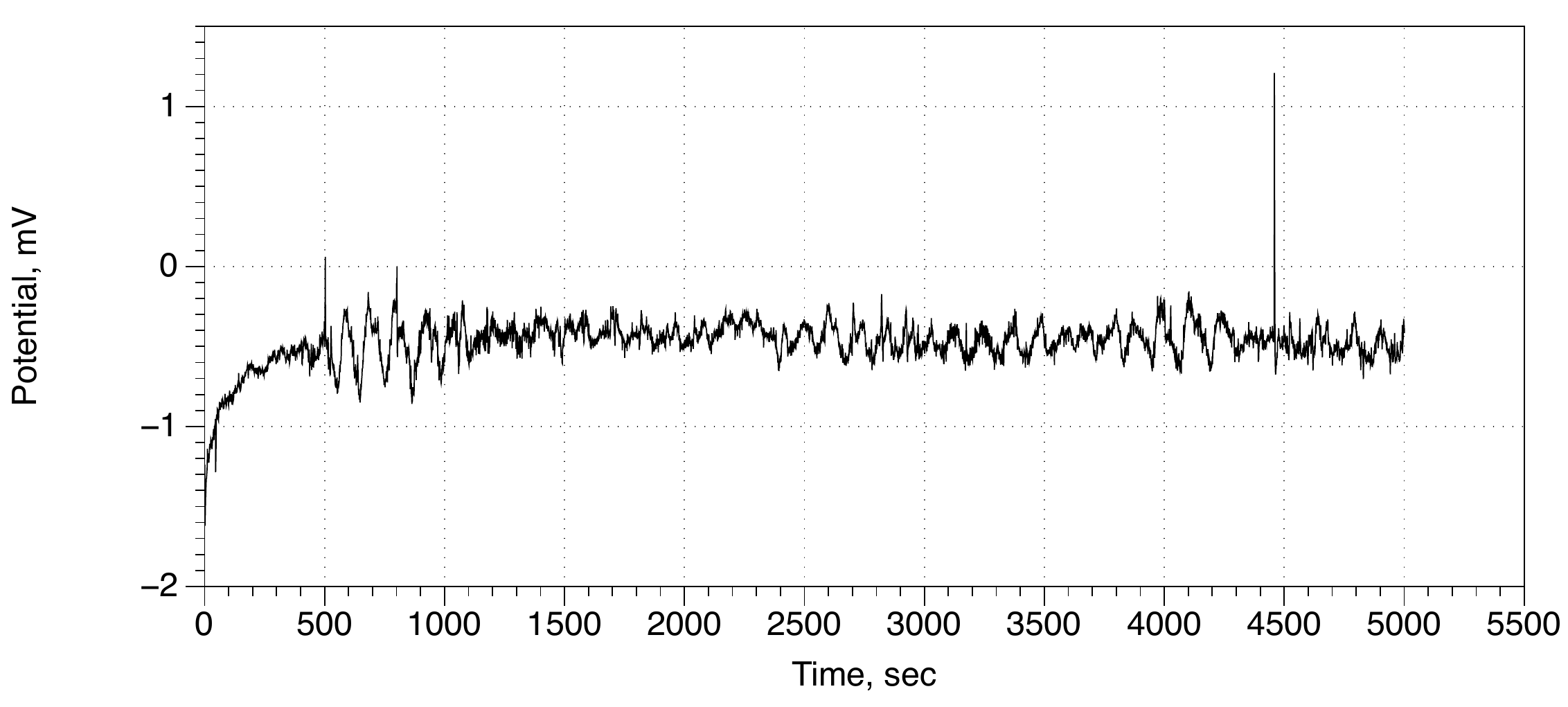,width=13cm}}
\vspace{8pt}
\caption{Trace to show differential voltage measurements from a typical control plasmodium
showing the first three phases of electrical potential activity, namely an initial depolarisation (initial drop not shown), followed by repolarisation and resting. Note the oscillations in the resting phase.}
\end{figure}

\subsubsection{Differential voltage measurements}
The most significant finding drawn from differential voltage measurements (Table 1) was
that the mean oscillating voltage (i.e. the voltage at which the plasmodium was observed to oscillate at in its 'resting' phase) was highly variable between treatments. This effect was most prominent in plasmodia treated with tungsten oxide: where control colonies only oscillated within a narrow band of voltages (up to 3mv), those treated with tungsten oxide would commonly oscillate at voltages up to 9mv. The effects of magnetite were similar, but not as pronounced (up to 6mv). It was also noted that in colonies treated with tungsten oxide NPs, the pre-oscillation (baseline) potential was frequently more polarised than that of controls (Fig. 12), which could have been a contributing factor towards the alteration of mean oscillating potential. There was also some variation in the mean wavelength/period of oscillation and mean amplitudes; there was great variation in all of the variables measured between samples, however.

\begin{table}
\tbl{Table to show collated experimental data from electrical testing of altered plasmodia (see text for heading descriptions).}
{\begin{tabular}{@{}cccc@{}} \toprule
 & Mean Wavelength (s) & Mean Amplitude (mv) & Mean Oscillating Vd \\ 
\colrule
Control\hphantom{00} & \hphantom{0}106.05 & \hphantom{0}0.55 & 0.47 \\
Tungsten Oxide\hphantom{00} & \hphantom{0}125.12 & \hphantom{0}0.61 & 3.30 \\
Magnetite\hphantom{0} & \hphantom{0} 100.60 & \hphantom{0}0.89 & 2.54 \\
\botrule
\end{tabular} \label{t1}}
\end{table}

\begin{figure}
\centerline{\psfig{file=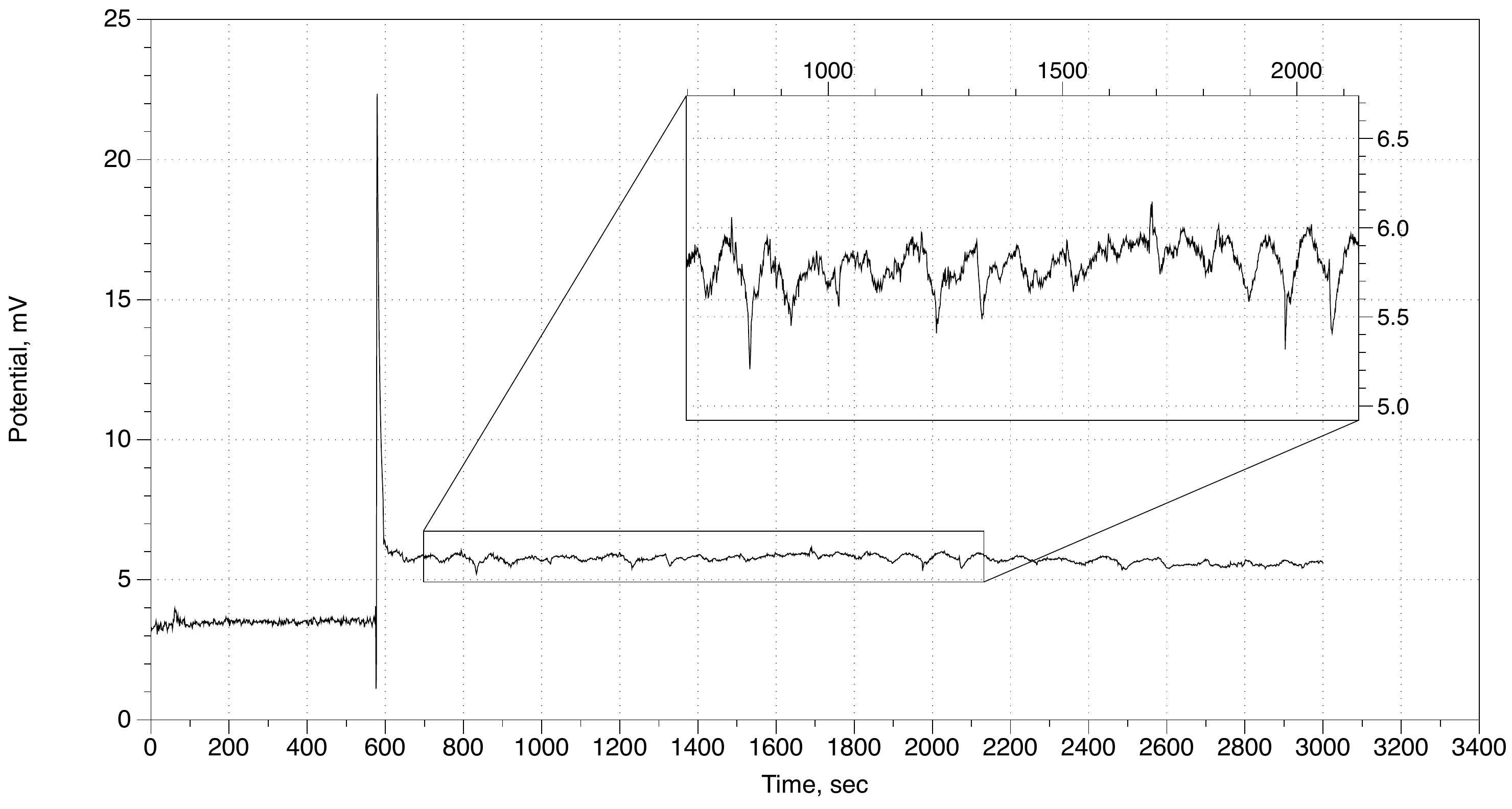,width=13cm}}
\vspace{8pt}
\caption{Graph to show hyperpolarisation (corresponding to colonisation of an agar blob at  c. 600 seconds) followed by oscillation in a plasmodium treated with tungsten oxide nanoparticles. Note how the baseline potential difference begins at ~3.5mv.}
\end{figure}

The waveforms of plasmodia treated with tungsten oxide NPs were also observably different
to those of the controls. In addition to being significantly more noisy with predilictions to
spontaneously alter their mean oscillating potential for no apparent reason, they were also prone to developing characteristic 'dips' in the apex (Fig. 13). These dipped apexes were discontinuous, occuring and resolving spontaneously, and did not reoccur in any discernable pattern.

\begin{figure}
\centerline{\psfig{file=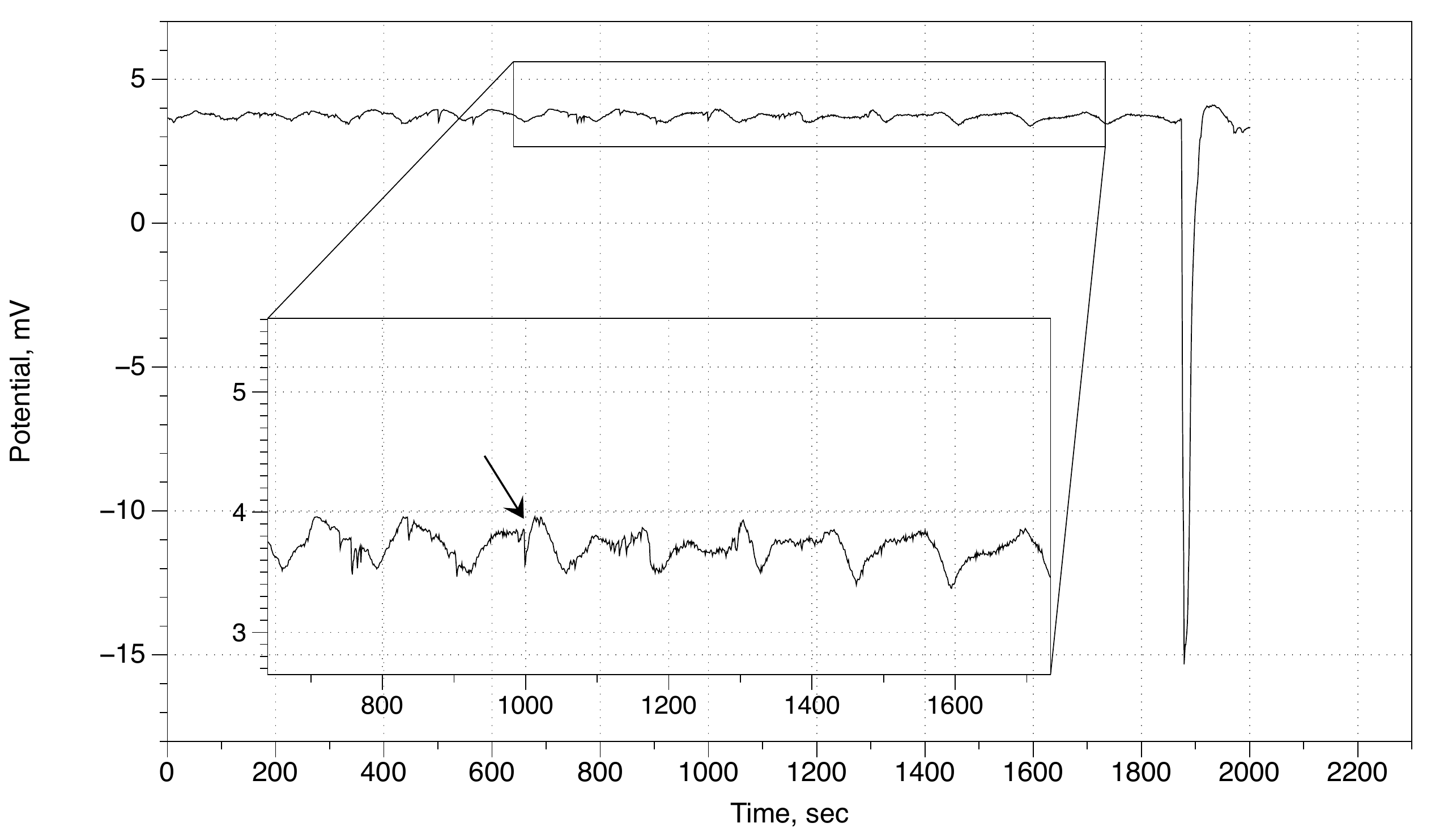,width=13cm}}
\vspace{8pt}
\caption{Oscillations in a plasmodium treated with tungsten oxide nanoparticles, showing the
characteristic 'dipped apex' (one of which is arrowed).}
\end{figure}

\subsubsection{Resistance measurements}
Electrical resistance in plasmodial tubes was found to oscillate, presumably in synchrony
with voltage oscillations and cytoplasmic streaming (Fig. 14). Resistance was measured in both
fully hydrated plasmodial tubes as well as desiccated, 'empty' tubes: the results of these
measurements are summarised in Table 2. The 'emptiness' of plasmodial tubes was assessed by eye rather than being measured after a specific period due to the variation in time it took for tubes to form.

It was found that plasmodia treated with tungsten oxide NPs had a higher average resistance
than controls when fully hydrated, although empty tubes were somewhat conductive in comparison
to controls which were effectively not conductive. Tubes from colonies treated with magnetite NPs were slightly more conductive than controls when fully hydated, but were considerably more so
when desiccated. The integrity of empty tube structures was occasionally comprimised by the
dessication of the agar blobs they were attached to; their resistance was therefore extremely high (500M$\Omega$ to infinity) and were not included in the dataset presented in table 2. It was also found that oscillating plasmodia would stop oscillating for long periods (2--6 hours) following resistance measurements.

\begin{figure}
\centerline{\psfig{file=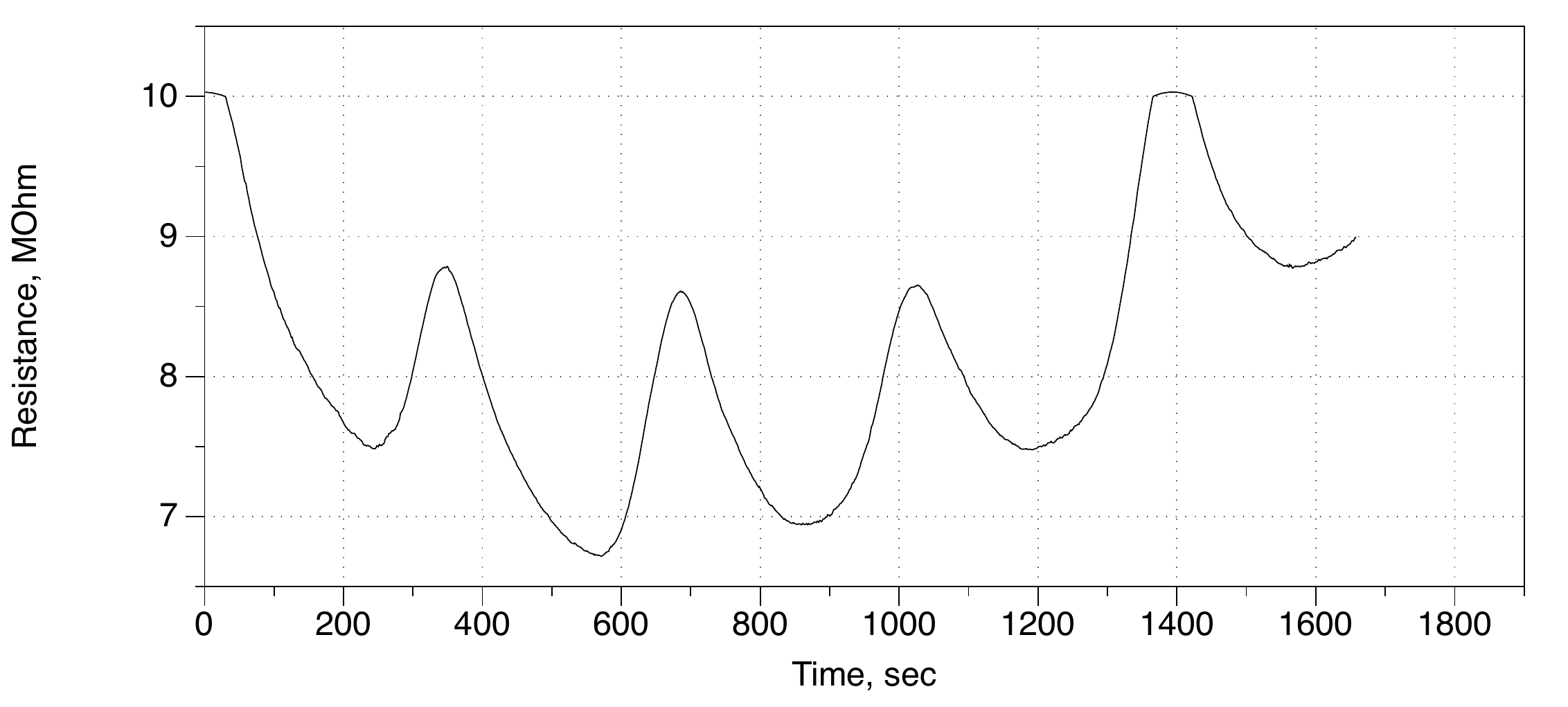,width=13cm}}
\vspace{8pt}
\caption{Graph to show oscillations in the resistance of an unaltered (control) plasmodial tube.}
\end{figure}

\begin{table}
\tbl{Table to show average resistance values for hydrated ('full') and desiccated ('empty')
plasmodial tubes.}
{\begin{tabular}{@{}cccc@{}} \toprule
 & Full Tube Mean Resistance & Empty Tube Mean Resistance \\ 
\colrule
Control\hphantom{00} & \hphantom{0}2.89M$\Omega$ & \hphantom{0}\textgreater500M$\Omega$ \\
Tungsten Oxide\hphantom{00} & \hphantom{0}4.70M$\Omega$ & \hphantom{0}63.50M$\Omega$ \\
Magnetite\hphantom{0} & \hphantom{0}1.42M$\Omega$ & \hphantom{0}0.88M$\Omega$ \\
\botrule
\end{tabular} \label{t1}}
\end{table}

\section{Conclusions}
\subsection{Assembly of nanoparticles within the \emph{P. polycephalum} plasmodium}
Imaging studies confirmed that the \emph{P. polycephalum} plasmodium is able to internalise
exogenous bio-compatible NPs, wherein they form a variety of structures which are transported
through the endoplasm before being deposited to the peripheral, non-motile regions of the
plasmodium. The observations made via confocal and electron microscopy provide strong evidence
for the occurrence of intraplasmodial assembly of both single and multiple NP varieties. With
reference to the aims of this investigation, the data presented here coupled with the observed
changes in electrical properties (see section 4.2) imply that \emph{P. polycephalum} can be successfully hybridised with metallic NPs.

Whilst intraplasmodial assembly was not obligatory (i.e. not all NPs underwent assembly), it
was nevertheless observed frequently, indicating that the generation of discrete electical devices within the plasmodium is a realistic possibility. The methodology presented here offers no control over the size, structure or localisation of these devices, however, and so any devices produced in this way hold no use as discrete devices; indeed, it was not discovered whether assembly occurred as a consequence of being absorbed into the plasmodium or as a function of their natural properties. More sophisticated methods of promoting binding between exogenous NPs, however, e.g. immunolocalisation, may prove to produce more uniform devices with distinct uses.

Although of little use for the generation of discrete devices, uncontrolled assembly of
metallic NPs was found to facilitate the generation of solid-state wires (see section 4.2); the mechanisms underlying this are likely to involve NPs depositing to the peripheral regions
following their exo- and/or transcytosis\cite{22} and forming a discontinuous layer which persists following migration away and/or sclerotinization. This results in a permanent plasmodially-assembled biomorphic mineralised network which facilitates the transmission of electricity. This network was demonstrated to be more conductive when not enrobed in living plasmodium when formed from magnetite NPs and \emph{vice versa} for tungsten oxide. The large difference in conductivities of tungsten oxide and magnetite trails demonstrates how the electrical properties of these networks may be artificially manipulated by altering NP type and presumably concentration. It is also possible that the conductive trail left in the wake of, but still connected to a moving plasmodium could be used as an interface between conventional hardware and the plasmodium.

The identification of small NP deposits 'floating' a few microns outside of the outer
membrane via TEM suggests that NPs are deposited in some quantity to the slime layer; this would further support the aforementioned hypothesis of how mineralised networks are formed by the plasmodium. An exhaustive literature search found no mention of external slime layer imaging with standard TEM, although in a study by Kuroda and Kuroda (see Ref. 21), calcium deposits are seen 'floating' a few microns away from the outer membrane, suggesting that the slime layer is still present following sample processing for electron microscopy, but is not stained by the procedure outlined in section 2.4. It was concluded that the NPs present in the slime layer were likely to be in the process of being endocytosed or are being exo- or transcytosed.

We propose that it would be advantageous to thoroughly explore NP localisation techniques
as the next step in the path to creating functional hybrid devices, e.g. immunolocalisation towards structural components such as actin filaments, or labelling NPs with proteins to encourage specific forms of assembly.

\subsection{Altering the electrical properties of {\emph P. polycephalum} with nanoparticles}
Loading metals into \emph{P. polycephalum} does change its electrical characteristics. Perhaps the
most useful electrical alteration that was explored in this investigation was that NPs can be used to increase the conductivity of both the live plasmodium and the empty desiccated tracks left in its wake (see section 4.1). The implications of self-growing – and presumably self-repairing – wires in the fields of biological circuit design and unconventional computing are manifold, as are the practical applications of a conductive intelligent organism, as discussed in the introduction. Although the differences in resistance/conductivity were distinct between treatments, there was great variation in treatment groups, highlighting that any technology that could be developed from this relatively simplistic methodology would not produce repeatable results.

Treatment with NPs was also found to alter the baseline potential and mean oscillating
potential of the plasmodium was frequently altered: in simpler terms, the differential voltage
between opposing poles of the plasmodium was significantly greater in altered plasmodia, both
when it was and was not oscillating. Whilst it is not unusual to observe differences in baseline potential in control colonies, which was \emph{circa} $\pm$ 1.5mv in the experiments run during this study, the oscillating potential would always return to $\pm$ 1.0mv. In contrast, both magnetite and tungsten oxide NP-treated colonies showed great variation in these variables: the former ranging from $\pm$ 8.4mv (mean 2.54) and the latter from $\pm$ 14.1mv (mean 3.30). The ability to create living devices with a pre-determined membrane potential is understandably an attractive prospect from a computing perspective. For such technologies to be developed, however, further work is needed to improve the repeatability of these variables: this is likely to be achieved by controlling the intake and intraplasmodial localisation of exogenous NPs as discussed in section 4.1.

It would also appear that the application of an external voltage source alters normal
plasmodial electrical activity, in that it consistently stopped oscillation for long periods of time. This was observed when oscillating plasmodia (observed via continuous differential voltage measurements) underwent resistance/conductance measurements before being re-connected to voltage-measuring apparatus. This was presumably a reaction to the insulting stimulus that was the application of a reference current. In all instances, the colony survived with no ill-effects from the external voltage source, but would take several hours to begin oscillating again. It would appear that, although \emph{P. polycephalum} displays galvanotaxis\cite{1}, it is sensitive to the application of relatively small sources of electrical potential. This could be used in further studies as a means of controlling the growth of the plasmodium, e.g. a plasmodium may be attracted from one electrode to the other via galvanotaxis, before it's electrical activity becomes stabilised (i.e. stops oscillating) by the electrical current that is passed down the resulting wire. Further experimentation was conducted following resistance measurements by applying a considerably larger current to the plasmodium: currents exceeding 1.5A (at 30v) seemed to dramatically reduce the resistance of the plasmodium from the order of megaohms to mere hundreds of ohms. Furthermore, the application of such a current (which could potentially kill a human if applied across the length of their body) was in no way detrimental to the health of the plasmodium, with all experimental colonies living for many days following exposure (n=5).

The alteration of oscillatory waveforms by the application of tungsten oxide NPs also
represents another means of increasing our control of plasmodial electrical activity. The cause of the dipped peaks observed (Fig. 13) are open to interpretation: the most likely explanation, however, is that the NPs interfere with the propagation of the calcium influx-induced waves which drive shuttle streaming via activation of muscle protein contraction. This likely causes the synchronous generation of two waves in close proximity to each other, resulting in two concurrent peaks of the same waveform. This could represent an interesting focus for study, as manipulation of waveform alters – in theory, at least – the information-processing ability of the plasmodium.

Most of the measurements made were subject to significant variation. Whilst this is only to
be expected when observing living systems, it is nevertheless unhelpful when trying to create a device with specific physical properties. This therefore further highlights the need to develop techniques to improve repeatability.

\subsection{Manipulating the behaviour of {\emph P. polycephalum} with nanoparticles}
Whilst attempts to load plasmodia with photoconductive zinc oxide NPs were unsuccessful,
by attempting to do so it was discovered that \emph{P. polycephalum} is highly intolerant to zinc oxide, and presumably many other zinc containing compounds due its biocidal nature in high concentrations. It is feasible, therefore, to suggest that zinc may be used as a chemorepellent substance in future studies; whilst many other compounds have been identified as Physarum repellents\cite{11}, substances with specific electrical properties that are also chemorepellents could potentially be of some use in the construction of hybrid Physarum-solid state arrays.

Altered plasmodia had a higher rate of failing to propagate between electrodes. The most
common response was for a plasmodium to colonise the blob overlying the reference electrode and remain viable for several days, before finally sclerotinizing. Sclerotinization seemed to correspond with the gradual desiccation of the agar, although occasionally the plasmodium would migrate in a different direction, e.g. up the aluminium tape. Whilst this also occurred in some of the control colonies, it was more frequent in those treated with tungsten oxide and even more so in those treated with magnetite NPs. These NPs were considered to be 'behaviour modifiers', however, as opposed to a chemorepellents or biocides, because they did not appear to be detrimental to the health of the plasmodium, as rapid negative chemotaxis, death or sclerotinisation were not observed in the test plasmodia with any greater frequency than in controls. In the case of magnetite, it is feasible that the plasmodium is able to sustain its self for some time on the hydrodynamic starch coating of the NPs, although it should be noted that a number of magnetite-treated colonies simply propagated in the wrong direction. This was observed on both the electrical testing apparatus and standard agar plates.

It is also feasible that the aqueous NP suspensions dispensed onto the reference electrode
agar blobs kept them more hydrated the blobs overlying the test electrodes, thereby giving the
plasmodium reason not to migrate away. It is also possible, however, that alterations in electrical properties may interfere with chemotactic mechanisms, e.g. by affecting the ionisation of proteins involved in membrane-bound receptor activation. This loss of chemotactic ability could therefore make programming altered plasmodia with chemical gradients problematic; any attempts to create ahybrid artificial-slime mould device with any use would have to overcome this.

\section{Acknowledgements}
The authors would like to extend the warmest thanks to:
\begin{itemlist}
\item Dr. David Patton, for his extensive collaboration, technical expertise with both varieties of electron microscope and aid in interpreting the imaging data presented.
\item Mr. David Corry, for his collaboration and technical expertise with confocal microscopy.
\item Mr. Paul Kendrick, for his histochemistry and light microscopy expertise.
\item Dr. Ben de Lacy Costello, for supervising the laboratory-based work.
\end{itemlist}
Figure 2 was created by the principal author using Inkscape Vector Graphics Editor, based on TEM observations and the descriptions of Stephenson and Stempen\cite{30}.

All photos were taken by the author using a Sony Xperia Ray smartphone. Light micrographs were taken using an Apex Minigrab camera attached to an Apex Practitioner light microscope.


\begin{thebibliography}{33}
\bibitem{1} A. Adamatzky, {\it Materials Science and Engineering} {\bf 30}, 1211 (2010).
\bibitem{2} A. Adamatzky, {\it Physarum Machines: Computers from Slime Mould},1st edn. (World Scientific Publishing, London, 2010).
\bibitem{3} A. Adamatzky and J. Jones, {\it Biophysical Reviews and Letters} {\bf 6}, 29 (2011).
\bibitem{4} A. Adamatzky, {\it IEEE Transactions on NanoBioscience} {\bf 11}, 131 (2012).
\bibitem{5} A. Adamatzky, {\it European Physics Journal E, Soft Matter} {\bf 31} 403 (2012).
\bibitem{6} A. Adamatzky, {\it ArXiv: 1306.0258}
\bibitem{7} O. Anderson, {\it Journal of Eukaryotic Microbiology} {\bf 40}, 67 (1993).
\bibitem{8} R. Beale and T. Jackson, {\it Neural Computing: An Introduction}, 1st edn. (IOP Publishing, Bristol, 1990).
\bibitem{9} V. Bonifaci, K. Mehlhorn and G Varma, {\it Journal of Theoretical Biology} {\bf 309}, 121 (2012).
\bibitem{10} L. Chua, {\it IEEE Transactions of Circuit Theory} {\bf 18}, 507 (1971).
\bibitem{11} B. Costello and A. Adamatzky, {\it Communicative and Integrative Biology} {\bf 6}, e25030 (2013). 
\bibitem{12} J. Fingerle and D. Gradmann, {\it Journal of Membrane Biology} {\bf 68}, 67 (1982).
\bibitem{13} J. Gaiarsa, O. Caillard and Y. Ben-Ari, {\it Trends in Neuroscience} {\bf 25}, 564 (2002).
\bibitem{14} E. Gale, A. Adamatzky and B. Costello, {\it ArXiv: 1306.2414}
\bibitem{15} K. Gerrow and A. Triller, {\it Current Opinions in Neurobiology} {\bf 20}, 631 (2010).
\bibitem{16} R. Ishikawa, T. Okagaki, S. Higashi-Fujime and K. Kohama, {\it Journal of Biological Chemistry} {\bf 266}, 21784 (1991).
\bibitem{17} N. Kamiya and S. Abe. {\it Journal of Colloid Science} {\bf 5}, 149 (1950).
\bibitem{18} H. Keller and S. Everhart, {\it Fungi} {\bf 3}, 15 (2010).
\bibitem{19} T. Kim, E. Jang, N. Lee, D. Choi, K. Lee, J. Jang, J. Choi, S. Moon and J. Cheon, {\it Nano Letters} {\bf 9}, 2229 (2009).
\bibitem{20} U. Kishimoto, {\it Journal of General Physiology} {\bf 41}, 1205 (1958).
\bibitem{21} R. Kuroda and H. Kuroda, {\it Journal of Cell Science} {\bf 44}, 75 (1980).
\bibitem{22} R. Mayne, D. Patton, B. Costello, A. Adamatzky and R. Patton, On the internalisation, intraplasmodial carriage and excretion of metallic nanoparticles in the slime mould {\it Physarum polycephalum}, to appear in {\it International Journal of Nanotechnology and Molecular Computation}
\bibitem{23} S. Mishra, R. Srivastava and S. Prakash, {\it Journal of Alloys and Compounds} {\bf 539}, 1 (2012).
\bibitem{24} T. Nakagaki {\it Research in Microbiology} {\bf 152}, 767 (2001).
\bibitem{25} H. Pang, Z. Li, X. Xiang, Y. Fu, F. Placido and X. Zu, {\it Applied Physics} {\bf 112}, 1033 (2013).
\bibitem{26} Y. Pershin, S. La Fontaine and M. Di Ventra, {\it Physical Review E} {\bf 80}, DOI: 021926 (2009).
\bibitem{27} Y. Pershin and M. Di Ventra, {\it Neural Networks} {\bf 23}, 881 (2010).
\bibitem{28} C. Reid, T. Latty, A. Dussutour and M. Beekman, {\it PNAS} {\bf 109}, 17490 (2012).
\bibitem{29} E. Ridgway and A. Durham, {\it Journal of Cell Biology} {\bf 69}, 223 (1976).
\bibitem{30} S. Stephenson and H. Stempen, {\it Myxomycetes: A Handbook of Slime Moulds}, 1st edn. (Timber Press, Oregon, 1994).
\bibitem{31} D. Strukov, G. Snider, D. Stewart and S. Williams, {\it Nature} {\bf 453}, 80 (2008).
\bibitem{32} K. Wohlfarth-Bottermann, {\it Journal of Experimental Biology} {\bf 67}, 49 (1977).
\bibitem{33} W. Wormington and R. Weaver, {\it PNAS} {\bf 73}, 3896 (1976).
\bibitem{34} S. Yoshihyama, M. Ishigami, A. Nakamura and K. Kazuhiro, {\it Cell Biology International} {\bf 34}, 35 (2010).

\end{thebibliography}
\end{document}